\documentclass[aps,prl,twocolumn,superscriptaddress]{revtex4}

\usepackage{graphicx}
\usepackage{amsmath}

\begin{document}

\title{Direct Experimental Evidence for the Hybridization of Organic Molecular Orbitals
with Substrate States at Interfaces: PTCDA on Silver}

\author{J. Ziroff}
\affiliation{Universit\"{a}t
W\"{u}rzburg, Experimentelle Physik VII, 97074 W\"{u}rzburg,
Germany}
\author{F. Forster}
\affiliation{Universit\"{a}t W\"{u}rzburg, Experimentelle Physik
VII, 97074 W\"{u}rzburg, Germany}
\author{A. Sch\"{o}ll}
\affiliation{Universit\"{a}t W\"{u}rzburg, Experimentelle Physik
VII, 97074 W\"{u}rzburg, Germany}

\author{P. Puschnig}
\affiliation{Chair of Atomistic Modelling and Design of Materials,
University of Leoben, 8700 Leoben, Austria}

\author{F. Reinert}
\affiliation{Universit\"{a}t W\"{u}rzburg, Experimentelle Physik
VII, 97074 W\"{u}rzburg, Germany}
\affiliation{Karlsruher Institut
f\"{u}r Technologie, Gemeinschaftslabor f\"{u}r Nanoanalytik, 76021
Karlsruhe, Germany}

\date{\today}

\begin{abstract}
We demonstrate the application of orbital $k$-space tomography for
the analysis of the bonding occurring at metal-organic interfaces.
Using angle-resolved photoelectron spectroscopy (ARPES), we probe
the spatial structure of the highest occupied molecular orbital
(HOMO) and the former lowest unoccupied molecular orbital (LUMO) of
one monolayer 3,4,9,10-perylene-tetracarboxylic-dianhydride (PTCDA)
on Ag(110) and (111) surfaces and in particular the influence of the
hybridization between the orbitals and the electronic states of the
substrate. We are able to quantify and localize the substrate
contribution to the LUMO and thus prove the metal-molecule hybrid
character of this complex state.
\end{abstract}

\pacs{33.60.+q, 73.20.Hb, 31.15.ae}

\maketitle

The adsorption of $\pi$-conjugated molecules on metals has attracted
considerable interest in surface science due to its importance for
the understanding of contacts in organic electronic devices. PTCDA
(3,4,9,10-perylene-tetracarboxylic-dianhydride) serves as an
archetypical molecule in this field
\cite{forrestCR1997,tautzPSS2007,glocklerSS1998}. In particular, its
behavior in molecular monolayers on Ag surfaces has been studied
intensely as a model system for chemisorptive molecule-metal
interaction \cite{glocklerSS1998,zouSS2006,bohringerSSL1998}. Rich
experimental data exist on the geometric structure of these
interfaces
\cite{bohringerSSL1998,braunSS2005,kilianPRL2008,hauschildPRL2005},
but the bonding distances differ quantitatively from theoretical
predictions \cite{rohlfingPRB2007,romanerNJoP2009}. Moreover,
spectroscopic techniques show a charge transfer from the metal into
the lowest unoccupied molecular orbital of the free molecule (LUMO),
but the underlying mechanism remains debated. Literature mentions
the involvement of several molecular orbitals \cite{zouSS2006}, Ag
$s$- and $d$-bands \cite{kawabeOE2008} and additional local bonds
forming between the carboxylic O atoms and the Ag surface
\cite{hauschildPRL2005}. Furthermore the Shockley surface state is
affected by the PTCDA adsorption and may play a role in the bonding
as well \cite{schwalbPRL2008,temirovNATURE2006}. A thorough
understanding of the interaction at the interface must be based on
the correct description of the valence orbitals. However,
information about the spatial distribution of molecular orbitals
(MOs) is not easily obtainable. So far scanning tunneling microscopy
(STM) is the most direct approach to image surface charge
distributions with sub-molecular resolution
\cite{rohlfingPRB2007,glocklerSS1998}. Another experimental method
is angle-resolved photoelectron spectroscopy (ARPES)
\cite{reinertNJoP2005,azumaJAP2000}, since the angular intensity
distribution pattern of organic molecules is linked to the wave
function geometry of the respective MOs
\cite{keraCP2006,keraJES2007}. However, interpretation of ARPES data
including final-state scattering \cite{osterwalderPRB1996} based on
atomic orbitals is challenging \cite{keraCP2006}. A new approach has
been described very recently \cite{puschnigSCIENCE2009} which
directly links photoemission intensity and initial state wave
function, enabling energy resolved tomographic imaging of the
electron distribution in $k$-space. This allows a direct comparison
of experimental and theoretical charge distribution and a probing of
the spatial structure of the bonding orbitals.

In this Letter, we present photoemission data of the two highest
occupied interface states of PTCDA on Ag(110) and Ag(111) surfaces
and compare our data to free-molecule orbitals as calculated by
density functional theory (DFT). We demonstrate the imaging of the
respective orbital structure and its correspondence to the
free-molecule HOMO/LUMO. Note here that we retain these terms also
for the respective orbitals of the adsorbed molecule for simplicity,
although the LUMO is not unoccupied anymore (while the HOMO becomes
the second occupied MO). Good quantitative agreement is achieved for
the spatial distribution of the (non-bonding) HOMO and DFT. The now
occupied LUMO deviates from its free-molecule form due to substrate
induced modifications. They take the form of an s-like contribution
to the orbital localized laterally at the center of the molecule.
Our interpretation of the electronic interaction is compared to
quantum-chemical calculations of PTCDA adsorbed on Ag(110)
\cite{abbasiJPC2009} and DFT results for PTCDA on Ag(111)
\cite{rohlfingPRB2007}. The experiments were performed in a UHV
setup composed of an organic molecular beam epitaxy chamber for
sample preparation and a spectrometer chamber for the ARPES
measurements (base pressure below 10$^{-10}$~mbar). The Ag
substrates have been prepared by standard sputtering and annealing
cycles as described elsewhere \cite{nicolayPRB2000}. PTCDA, purified
by triple-sublimation, was evaporated from a Knudsen cell at rates
of approximately 0.2~ML per minute onto the substrate which was kept
at room temperature. The monolayer coverage was controlled by
monitoring the continuous quenching of the Shockley type surface
state \cite{ziroffSS2009} and the characteristic PTCDA valence band
spectra of the chemisorbed monolayer \cite{zouSS2006}. All ARPES
measurements were performed at sample temperatures T~$\approx$~80~K
with a high-resolution photoelectron analyzer (Scienta R4000) in
combination with a monochromatized VUV lamp using He~I$_{\alpha}$
radiation ($h\nu$~=~21.23~eV) at an energy resolution of
$\Delta$E~=~5~meV. The angle resolved mode of the analyzer covers a
parallel detection range of up to $\pm~15^\circ$ with a resolution
of $\approx~0.3^\circ$. An additional rotation of the sample was
used to allow 2D $k$-space mapping. Molecular orbitals of an
isolated PTCDA molecule were calculated within the framework of DFT
using the ABINIT software package \cite{gonzeCMS2002}.
Norm-conserving pseudo-potentials with a cut-off of 50 Ryd and a
generalized gradient approximation for the exchange-correlation
energy and potential have been used. The calculated 2D-ARPES
intensity maps are obtained from the Fourier transform of the HOMO
and LUMO orbitals at the appropriate kinetic electron energies.
Details of this approach are described elsewhere
\cite{puschnigSCIENCE2009}.

 \begin{figure}
 \includegraphics[width=9.1cm]{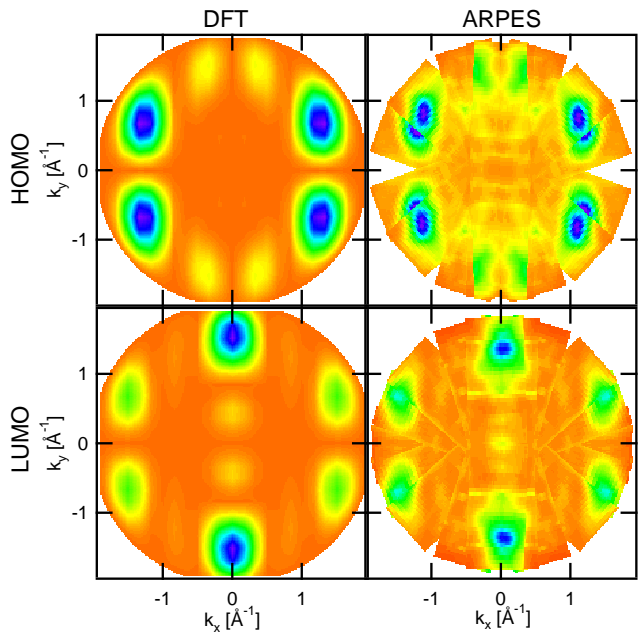}
 \caption {Right column: Experimental $k_{x,y}$-dependent ARPES
intensity of the HOMO (top) and LUMO (bottom) of PTCDA on Ag(110)
recorded with HeI$_{\alpha}$. Left column: Corresponding DFT
calculations of the free molecule at the respective photoelectron
energies.}
 \end{figure}

The monolayer of PTCDA on Ag(110) is a single-domain structure with
one molecule per unit cell. The identity of all molecules in terms
of adsorption site and absolute orientation allows the effective
mapping of MOs without the superposition of additional orientations.
The molecule grows on Ag(110) oriented along the [001]-axis of the
substrate \cite{seidelSS1997}. We will refer to this direction in
momentum space as $k_{x}$, to [1\=10] as $k_{y}$. Fig.~1 compares
the experimental ARPES intensity (right column) recorded at binding
energies of 0.8~eV and 1.9~eV with respect to the Fermi level to the
corresponding calculations for the valence MOs (left column). The
shown data set was completed from measurements covering one quadrant
using the symmetry axis of the system \cite{seidelSS1997}. From the
overall similarity we conclude that the measured orbitals correspond
to the HOMO and the LUMO respectively. However, the experimental
data exhibits additional photoemission intensity features which do
not appear in theory. They are mainly caused by the $sp$-bands of
the Ag substrate.
\begin{figure}
 \includegraphics[width=8.6cm]{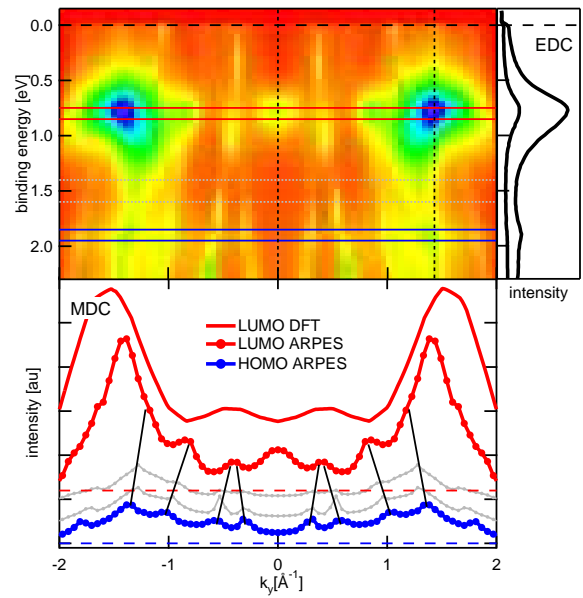}
 \caption{Top: ARPES intensity vs. $k_{y}$ at $k_{x}$~=~0 for 1 ML PTCDA/Ag(110),
EDC for indicated $k_{y}$ on the right. Bottom: MDC deduced from the
MOs energy windows (red/blue dotted line, dashed line marks
respective zero) and at intermediate energies (gray curves). The
black lines indicate the dispersing substrate bands. The solid red
curve shows the theoretical LUMO intensity (offset from zero for
better comparison).}
 \end{figure} Fig.~2 (top) shows an energy-resolved ARPES spectrum along the
$k_{y}$-direction at $k_{x}$~=~0 through the momentum map in Fig.~1.
The data allows a clear distinction between photoemission intensity
from the MOs and the strongly dispersing substrate bands. Selected
energy distribution curves (EDC) are displayed on the right. Fig.~2
(bottom) depicts the energy-integrated $k_{y}$-dependent
photoemission intensity of the LUMO (red) and the HOMO (blue).
Additional line scans (momentum distribution curves (MDC)) at
intermediate binding energies (gray lines) are added to visualize
the contribution of the Ag bands, black lines guide the eye along
the dispersion. The momentum distribution of the HOMO signal is
basically featureless, in agreement with theory (cf. Fig.~1). The
LUMO signal also generally resembles the free-molecule calculations.
However, there is an additional intensity maximum at normal emission
($k_{||}$~=~0). Note that this is neither present in the free
molecule nor accidentally caused by substrate bands. In addition,
the position of the dominating emission maxima at
$k_{y}$~$\approx$~1.6~\AA$^{-1}$ is shifted to smaller absolute
$k_{y}$-values. Interpreting our ARPES results for PTCDA on Ag(110),
we find the HOMO generally unaltered by the adsorption, therefore we
rule out significant substrate admixture to this orbital, the HOMO
is apparently not involved in the bonding at the interface. The LUMO
generally resembles the free-molecular calculations. However, the
shift of the characteristic main intensity maxima to lower $k_{y}$
values indicates a distortion of the orbital structure in this
direction (i.e. the short molecular axis). Moreover, additional
intensity appears around $k_{||}$~=~0. In real space this would
correspond to a laterally node-free charge distribution at the
center of the molecule. We therefore suggest a bonding of the
molecule due to LUMO hybridization, taking the form of an s-like
substrate contribution to the MO at the PTCDA perylene core. A
comparison with recent quantum-chemical calculations reconfirms the
observed negligible interaction of the HOMO \cite{abbasiJPC2009}.
For the LUMO, a distortion along the short molecular axis and
additional substrate-localized lobes are visible in the calculated
orbital structure \cite{abbasiJPC2009}, matching our experimental
findings. However, the calculations show a node (and therefore a
parity change) of the hybrid orbital along the $k_{x}$ axis, a
prediction which is not compatible with the experimentally observed
normal emission intensity.

\begin{figure}
 \includegraphics[width=8.6cm]{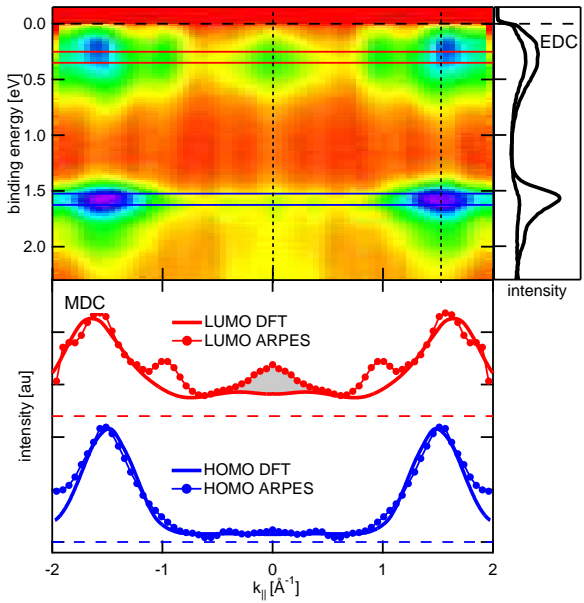}
 \caption{Top: ARPES intensity vs $k_{||}$ for 1~ML PTCDA/Ag(111),
EDC for indicated $k_{||}$ on the right. Bottom: MDC of the LUMO
(red curve) and HOMO (blue curve) intensity, respective energy
windows marked above, zero indicated by dashed line. The solid lines
correspond to the calculated DFT intensities (offset from zero).
Gray shaded area models the normal emission intensity as Gaussian
(FWHM~=~0.42~\AA$^{-1}$)}
 \end{figure}

In contrast to the (110) surface, PTCDA on Ag(111) forms a
multi-domain herringbone structure. Six symmetry equivalent domains
with two molecules per unit cell \cite{glocklerSS1998} result in
twelve different molecule orientations. Therefore, measuring the
polar-angle dependence of the PE intensity effectively means probing
the averaged ${k_{||}}$-dependence of the MOs. While this reduces
the 2D-maps into one-dimensional data sets it gives two advantages:
On the one hand a more complex geometry of the organic overlayer
prevents an overlap with intense $sp$-bands due to backfolding
\cite{ziroffSS2009}. On the other hand one measures all data points
at an identical geometry between photon source and the emitted
electrons, allowing for a strict quantitative analysis of the MDC in
terms of electron distribution of the initial MO, since the
modulating polarization factor $\textbf{A}~\cdot~\textbf{k}$ remains
constant \cite{puschnigSCIENCE2009}. Fig.~3 presents energy and
$k_{||}$ resolved photoemission intensity for a monolayer of
PTCDA/Ag(111) (top, data symmetrized to $k_{||}$~=~0). The MDCs of
the background corrected LUMO and HOMO (red and blue energy window)
are compared to the free-molecule DFT results in the bottom panel.
The theoretical data was averaged over the azimuthal angle
(emulating the multiple domains in the experiment) and offset from
zero to allow for a constant experimental intensity.  The HOMO is
again characterized by a single, dominating maximum. The deviation
for higher $k$-values is probably caused by the grazing incidence of
the light and high electron emission angle
($\Theta$~$\approx$~70$^\circ$), setting an upper $k$ limit for
quantitatively reliable measurements in our experimental setup.
Otherwise, the agreement between free-molecule theory and experiment
is excellent. From this we conclude an undistorted HOMO geometry and
a weakly interacting character of this MO on Ag(111), analogous to
the Ag(110) substrate. For the LUMO, a slight inward shift of the
dominating maxima is observed, as is again ARPES intensity at
$k_{||}$~=~0. From this we deduce a similar bonding mechanism as
described for Ag(110) above. There is, however, also a distinctive
new peak at $k_{||}$~$\approx$~$\pm$~0.9~\AA$^{-1}$, neither present
in the DFT calculations nor a feature of the substrate. It
corresponds to a real-space periodicity of 7~\AA~in the LUMO
structure. This value is almost twice as large as the typical
node-distance within the undistorted MO (causing the dominating
ARPES intensity maxima) and comparable to the lateral PTCDA
dimensions (9.2~$\times$~14.2~\AA) \cite{glocklerSS1998}. This
length scale hints at a charge redistribution over the single PTCDA
molecule, possibly due to inter-molecular interaction. On Ag(111)
PTCDA arranges in a herringbone structure with a nearest-neighbor
distance below the van der Waals radii of the molecules, caused by
electrostatic forces acting between the negatively charged anhydride
group and the aromatic core region \cite{glocklerSS1998}. Such an
interaction could cause the observed modifications in a strongly
delocalized bonding orbital like the LUMO. Another indication is the
absence of the considered ARPES feature in the previously discussed
data for the Ag(110) surface, on which the PTCDA molecules are
distinctly separated from each other on specific adsorption sites.
In order to quantitatively analyze the additional normal-emission
intensity, we approximate it by a Gaussian as displayed by the
shaded area in Fig.~3. The real-space correspondence is a node-free
charge spread laterally (FWHM~$\approx$~~7~\AA) over the central
part of the PTCDA perylene core. A first estimate assigns
$\approx$~10~\% of the total PE intensity to the area around
$k_{||}$~=~0, a value which may provide a measure for the substrate
contribution to the PTCDA-Ag hybrid state. Quantum-chemical
calculations for PTCDA/Ag(110) state a comparable (16~\%) substrate
contribution \cite{abbasiJPC2009}. A detailed comparison with
theoretical calculations for PTCDA on Ag(111) is not directly
possible, there are only projected charge densities for
PTCDA/Ag(111) obtained by DFT calculations \cite{rohlfingPRB2007}.
They describe the lateral distribution between the molecule and the
substrate as an undistorted LUMO with additional charge accumulated
around the central carbon ring of the molecule \cite{tautzPSS2007}.
This description resembles the electron probability density that
would result from the orbital geometries deduced from our data.

In conclusion, we demonstrate the potential of orbital tomography by
ARPES for the comprehensive understanding of the fundamentals of the
bonding of large organic molecules to surfaces. On the particular
example of the PTCDA molecule, chemisorbed on Ag surfaces, we show
that the HOMO orbital does not show a significant substrate
admixture. It is therefore not decisively involved in the
molecule-substrate bonding. In contrast, the LUMO features a
significant distortion of the orbital spatial structure compared to
the free-molecule and a prominent admixture of substrate states. A
first analysis estimates the substrate contribution to the LUMO as
$\approx$~10\%, localized in a node-free charge accumulation at the
central carbon ring of the perylene core. Our results are in good
agreement with quantum-chemical calculations for the MOs of PTCDA on
Ag(110) and DFT calculations for PTCDA on Ag(111), with the
exception of the nodal structure of the LUMO hybridization on
Ag(110).

\begin{acknowledgments}
This work was supported by the Bundesministerium f\"{u}r Bildung und
Forschung (BMBF) under grants 05KS7FK2/05KS7WW1 and 035F0356B and by
the Deutsche Forschungsgemeinschaft (DFG) via FOR1162. P.P.
acknowledges  support from the Austrian Science Foundation (FWF)
within the national research network S97.
\end{acknowledgments}

\end{document}